\documentclass[twocolumn,showpacs,amsmath,amssymb]{revtex4}
\def \be {\begin{equation}}
\def \ee {\end{equation}}
\def \bea {\begin{eqnarray}}
\def \eea {\end{eqnarray}}
\def \nn {\nonumber}
\newcommand{\eq}[1]{(\ref{#1})}

\def\d{\delta}

\def\l{\lambda} 
\def\m{\mu} \def\n{\nu}

\def \d {\delta}

\def \m {\mu}
\def \n {\nu}

\newcommand{\hh}{\, ,\hspace{0.5cm}}
\newcommand{\hhh}{\, ,\hspace{0.2cm}}

\newcommand{\BM}[1]{{\mbox{\boldmath $#1$}}}

\newcommand{\BFE}[1]{\mbox{\bf e}_{\hat{#1}}}

\begin{document}
\title{String Gyratons in Supergravity}
\author{Valeri P. Frolov$^{1,2}$}  
\email{frolov@phys.ualberta.ca}
\author{Feng-Li Lin$^{3}$}
\email{linfengli@phy.ntnu.edu.tw}
\affiliation{$^{1}$ Theoretical Physics Institute, University of Alberta,
Edmonton, Alberta, Canada, T6G 2J1}
\affiliation{$^{2}$ Kavli Institute for Theoretical Physics,
University of California
Santa Barbara, CA 93106-4030}
\affiliation{$^{3}$ Department of Physics, National
Taiwan Normal University, Taipei City,
116, Taiwan}
\date{\today}

\begin{abstract}
We study solutions of the supergravity equations with the
string-like sources moving with the speed of light. An exact
solution is obtained for the gravitational field of a boosted ring
string in any dimension greater than three.
\end{abstract}

\pacs{04.70.Bw, 04.50.+h, 04.20.Jb \hfill
NSF-KITP-06-13, \ \ \
Alberta-Thy-02-06}


\maketitle


\section{Introduction}

In this paper we study solutions of supergravity equations generated
by string-like sources moving with the speed of light. Study of
solutions of the Einstein equations for objects moving with the
velocity of light has long story. In 1934 Tolman \cite{To} obtained
solutions for the gravitational field of beams of light in the linear
approximation. The exact solutions of the non-linear Einstein
equations for this problem were obtained later \cite{Pe1,Pe2,Bo1}.
For the infinitely small cross-section of the beam and for the
delta-type distribution of the light-pulse in time, these solutions
reduce to the Aichelburg-Sexl metric \cite{AiSe}. These 4-dimensional
solutions were generalized to the case the matter of beam is either
charged or spinning by Bonnor \cite{Bo2,Bo3} in 1970.
Higher-dimensional generalizations of the solutions of Einstein
equations for spinning relativistic beam-like sources (gyratons) were
obtained and studied in \cite{FrFu:05,FrIsZe:05}. Solutions for
charged higher-dimensional gyratons and for gyratons moving in an
asymptotically anti-de Sitter space were found in \cite{FrZe:05a} and
\cite{FrZe:05b}, respectively. The aim of this paper is to discuss
generalization of electrically charged gyraton solutions to the
theory of supergravity. More concretely, we study solutions of
higher-dimensional gravitational equations with the rank 3
antisymmetric Kalb-Ramon field generated by string-like sources
moving with the velocity of light. The basic equations, the metric
and field ansatz are presented in sections~2 and 3, respectively. In
section~4 we demonstrate, that the field equations for the problem
under consideration reduce to the set of linear equations in a flat
Euclidean space. Special solutions of these equations for rink-like
string gyratons are obtained in section~5. Section~6 contains brief
summary and discussions.

\section{Basic Equations}

We consider the massless bosonic sector of supergravity. We restrict
ourselves by discussing what is called the common sector. The fields
in the common sector are the metric $g_{\mu\nu}$, the Kalb-Ramond
antisymmetric field $B_{\mu\nu}$ and the dilaton field $\phi$. The
corresponding action, which is also the low-energy superstring
effective action, is
\[
S={1\over 16\pi G} \int d^D x \, \sqrt{|g|} e^{-2\phi} [ R -4(\nabla
\phi)^2
\]
\be\label{action} -{1\over 12} H_{\mu\nu\lambda}H^{\mu\nu\lambda}
]+{1\over 2}\int d^D x \, \sqrt{|g|} B_{\mu\nu}J^{\mu\nu}+{\cal
S}_{m}\, . \ee Here $G$ is the $D-$dimensional gravitational
(Newtonian) coupling constant, and  ${\cal S}_{m}$ is the action for
the string matter source.  The string coupling constant $g_s$ is
determined by the vacuum expectation value of the dilaton field
$\phi_0$, $g_s=\exp (\phi_0)$. The 3-form flux \be\label{H}
H_{\mu\nu\lambda}=
\partial_{\mu}B_{\nu\lambda}+\partial_{\nu}B_{\lambda\nu}
+\partial_{\lambda}B_{\mu\nu}
\ee
is the Kalb-Ramond (KR) field strength and $B_{\mu\nu}$ is its
anti-symmetric 2-form potential.
The field $H_{\mu\nu\lambda}$
is invariant under the gauge transformation
\be
B_{\m\n} \rightarrow B_{\m\n}+\partial_{\m} \Lambda_{\n}-\partial_{\n} \Lambda_{\m}.
\ee
$J^{\mu\nu}$ is the antisymmetric tensor of the current which plays
the role of a source for the KR field. For example, for the
interaction of the KR field with a fundamental string described by
the action
\be\label{int}
S_{int}=-{q\over 2} \int \d^2 \zeta \epsilon^{ab} B_{\mu\nu} {\partial X^{\mu}\over
\partial \zeta^a} {\partial X^{\nu}\over \partial \zeta^b}
\ee
this current is
\be
J^{\m\n}(x)={q\over 2}\int  \d^2 \zeta {\delta^D(x-X(\zeta))\over
\sqrt{|g|}} \epsilon^{ab}{\partial X^{\mu}\over
\partial \zeta^a} {\partial X^{\nu}\over \partial \zeta^b}\, .
\ee
Here $\epsilon^{ab}$ is the antisymmetric symbol,
$\zeta^a=(\tau,\sigma)$ are parameters on the string surface and the
functions $X^{\mu}=X^{\mu}(\zeta)$ determine the embedding of the
string worldsheet in the bulk (target) spacetime. The parameter $q$
is the "string charge". The current $J^{\m\n}$ is tangent to the
worldsheet of the string, $J^{[\m\n} X^{\lambda ]}_{,c}=0$.

We shall study a special class of gyraton solutions for which the
dilaton field is constant, i.e., $e^\phi=g_s$. In this case the
field equations  are
\bea\label{5} && R_{\m\n}-{1\over2} g_{\m\n} R=
T_{\m\n}+g_s^2 \kappa {\cal T}_{\mu\nu}\,
, \\
&& H_{\m\n\;\;\; ; \l}^{\;\;\;\l}=8 \kappa J_{\m\n}. \label{6}
\label{Heq}
\eea
Here the stress-energy tensor for the 3-form flux is
\be\label{Tmn}
T_{\m\n}={1\over12} (3 H_{\m\l\rho}H_{\n}^{\;\;\l\rho}
-{1\over2} g_{\m\n}H_{\rho\sigma\l}H^{\rho\sigma\l})\, ,
\ee
and  $\kappa =8\pi g_s^2 G$. ${\cal T}_{\mu\nu}$ which enters the
equation (\ref{5}) is the stress-energy of the matter (string) which
we shall specify later.

Let $\Sigma$ be a $(D-2)-$dimensional spacelike surface, and
$\partial \Sigma$ be its boundary. We define the charge of the
fundamental string intersecting $\Sigma$ by Gauss's law as
\be
Q:=\int_{\partial \Sigma} d\sigma_{D-3}\; *_D H_3
=\int d\sigma_{\mu\nu\lambda}H_{\mu\nu\lambda}\, .
\ee
By using the Stoke's theorem  and (\ref{Heq}) one has
\be\label{charge}
Q=\int d\sigma_{\mu\nu}J^{\m\n}= 8 \kappa q\, .
\ee
Here $d\sigma_{\mu_1\cdots \mu_n}:=i_{\mu_1}\cdots i_{\mu_n} (*1)$ in which $*1$ is the volume form of $D$-dimensional spacetime and $i_{\mu}:\Lambda^p T^*\rightarrow \Lambda^{p-1}T^*$, $i_{\mu}dx^{\nu_1}\wedge \cdots \wedge dx^{\nu_p}=p\delta_{\mu}^{[\nu_1} dx^{\nu_2}\wedge \cdots \wedge dx^{\nu_p]}$.

For a straight string along $Z-$axis in Minkowski spacetime with
coordinates $(T,Z,X^i)$, (i=3,...,D) one has
\be
H^{T Z i}= {Q \over {\cal A}_{D-3}} {n^i\over
[\sum_{i=3}^{D} X_i^2]^{(D-3)/2}}\, ,
\ee
the other components vanish. Here $n^i$ is a unit vector normal to
the surface $\sum_{i=3}^{D} X_i^2=$const, and
\be\label{ss}
{\cal A}_n=2\pi^{n/2}/\Gamma(n/2)
\ee
 is the surface area of a unit
$n-$dimensional sphere.

In what follows, we consider the gravitational and KR fields outside
the sources, that is in the region where $T_{\m\n}=0$ and
$J_{\m\n}=0$. The relation (\ref{charge}) will be used to relate the
parameters which enter a solution to the charge of the string.

\section{Ansatz for Supergravity Gyraton}

\subsection{Metric}

We shall study special solutions of the sypergravity equations
(\ref{5})-(\ref{6}) which are generated by sources moving with the
speed of light. We are interested in solutions which have finite
energy, angular momentum, KR charge, and finite duration in time.
Following \cite{FrFu:05,FrIsZe:05,FrZe:05a} we call such
ultrarelativistic objects with spin {\em gyratons} and use for
its gravitational field  in the $D=n+2$ dimensional spacetime
the following metric ansatz (Brinkmann metric \cite{Brinkmann:25})
\be\label{mansatz}
ds^2=d\bar{s}^2+2(a_u du+ a_a dx^a)du.
\ee
Here $a_u=a_{u}(u,x^a)$, $a_a=a_{a}(u,x^a)$,
and
\be\label{fmetric}
d\bar{s}^2=-2dudv+d\BM{x}^2
\ee
is the D-dimensional flat metric, the transverse spatial part of the
metric $d\BM{x}^2=\delta_{a b} dx^a dx^b$ in the n-dimensional
hypersurface is flat. We use the Greek letters for indices which take
values $1,\cdots,D$, while the Roman low-case indices take value
$3,\cdots, D$. The form of the metric (\ref{mansatz}) implies that
\be\label{det}
\det(g_{\m\n})=-1\, .
\ee

The field
\be
a_{\mu}=a_{u} \delta^u_{\mu}+a_{a} \delta^a_{\mu}
\ee
is the gravitational analogue of the electromagnetic potential. It is
easy to see that the metric is invariant under the following gauge
transformation
\be\label{gaugg}
v \rightarrow v+\lambda(u,\BM{x}), \qquad a_{\mu} \rightarrow a_{\mu}
-\lambda_{,\mu}\, ,
\ee
and the quantity
\be
f_{\mu\nu}=\partial_{\mu} a_{\nu}-\partial_{\nu} a_{\mu}
\ee
is gauge invariant.

The metric (\ref{mansatz}) admits a null Killing vector $l=l^{\mu}
\partial_{\mu}=\partial_v$, which is parallelly propagated
$l_{\mu;\nu}=0$. One also has
\be
l_{\mu}dx^{\mu}=-du\hh l^{\mu}a_{\mu}=l^{\mu}f_{\mu\nu}=0\, .
\ee

The flat metric $d\bar{s}^2$ and the metric (\ref{mansatz}) are
related as
\bea\label{gg1}
&&\bar{g}_{\mu\nu}=g_{\nu\nu} -l_{\mu} a_{\nu} -l_{\nu} a_{\mu} \, ,\\
&&\label{gg2} \bar{g}^{\mu\nu}=g^{\mu \nu}+l^{\mu} a^{\nu}+l^{\nu} a^{\mu}
+l^{\mu}l^{\nu} a^{\rho}a_{\rho}\, .
\eea

\subsection{KR field}

For the KR field potential we use the ansatz similar to the one
adopted for the electromagnetic gyratons \cite{FrZe:05a}. Namely
we postulate that $B_{\mu\nu}=B_{\mu\nu}(u,\BM{x})$ and
\be\label{cons1}
l^{\mu}B_{\mu\nu}=0\, .
\ee
It is easy to check that
\be\label{cons}
l^{\mu}H_{\mu\nu\lambda}=0.
\ee
The imposed constraints imply that the only non-vanishing components
of $B_{\mu\nu}$ are $B_{ua}(u,\BM{x})$ and $B_{ab}(u,\BM{x})$, and
of $H_{\m\n\l}$ are $H_{uab}(u,\BM{x})$ and $H_{abc}(u,\BM{x})$.
Moreover, to preserve the constraint (\ref{cons1}) under gauge
transformation, we should impose
\be
\partial_v \Lambda_{\m}-\partial_{\m}\Lambda_v=0.
\ee

  Using \eq{gg1}-\eq{gg2} it is easy to derive the relation between the contravariant tensors raising their indices from the same covariant one by using the metric \eq{mansatz} and the flat metric \eq{fmetric} respectively. Especially, for 3-form flux we have  
\bea
\bar{H}_{\m\n}^{\;\;\;\l}&=&H_{\m\nu}^{\;\;\;\l}+
l^{\l}a_{\rho} H_{\m\n}^{\;\;\;\rho}\, ,\\
\bar{H}_{\m}^{\;\;\n\l}&=&H_{\m}^{\;\;\n\l}+
a_{\rho}(l^{\lambda}H_{\m}^{\;\;\n\rho}-l^{\n}H_{\m}^{\;\;\l\rho})\, ,\\
\bar{H}^{\m\n\l}&=&H^{\m\n\l} \nonumber  \\
&+&a_{\rho}(l^{\l}H^{\m\n\rho}
-l^{\n}H^{\m\l\rho}+l^{\m}H^{\n\l\rho})\, .
\eea
Here the quantities with bar are the ones with respect to the flat metric.

One also has
\bea
&&\bar{H}_{\m\l\rho}\bar{H}_{\n}^{\;\;\l\rho}=H_{\m\l\rho}H_{\n}^{\;\;\l\rho}\,
,\\
&&H^2\equiv \bar{H}_{\m\n\l}\bar{H}^{\m\n\l}=H_{\m\n\l}H^{\m\n\l}\, .
\eea
These equations will be useful in solving the field equations. 

Moreover, the constraint (\ref{cons}) implies that
\be
H^2=\BM{H}^2\equiv H_{abc}H^{abc}.
\ee
Note that the indices $a,b,c$ in the above relations are raised by
the flat metric $\delta^{ab}$ since $g^{ua}=\bar{g}^{ua}=0$ and
$g^{ab}=\bar{g}^{ab}=\delta^{ab}$.

\subsection{Gyrating matter}

We discuss now the ansatz for ${\cal T}_{\mu\nu}$ which enters the
equation (\ref{6}). We require that this tensor obeys the conservation
low
\be
{\cal T}^{\mu\nu}_{\ \ ;\nu}=0\, ,
\ee
and is aligned to the null Killing vector $l_{\mu}$
\be
{\cal T}_{\mu\nu}=l_{(\mu}p_{\nu)}\hh
l^{\mu}p_{\mu}=0\, .
\ee
The last condition guarantees that the trace of ${\cal T}^{\mu\nu}$
vanishes, ${\cal T}^{\mu}_{\ \mu}=0$.
For the metric (\ref{mansatz}) these conditions are satisfied when
\be
p_{\mu}=p_{\mu}(u,x^a)\hh
p^a_{\ ,a}=0\, .
\ee
This can be checked by using the condition $l_{\mu;\nu}=0$. Bonnor
\cite{Bo3} called such matter in 4-dimensional spacetime spinning
null fluid.

\section{Reduced Equations}

Calculations show \cite{FrIsZe:05} that for the metric
(\ref{mansatz})  the only non-vanishing components of the Ricci
tensor are
\bea
&&R_{ua}={1\over2}f_{ab}^{\;\;\;,b}\, ,\\
&&R_{uu}=-(a_u)_{,a}^{,a}+{1\over4}f_{ab}f^{ab}+\partial_u(a_a^{\;,a})\,
.
\eea
These relations imply
\be
R=0, 
\ee
which together with the fact ${\cal T}^{\mu}_{\mu}=0$ the field equation \eq{5} yields 
\be\label{tr}
T^{\m}_{\m}=0\, .
\ee

   By taking trace of (\ref{Tmn}) the vanishing trace of the stress tensor then implies  
\be
\BM{H}^2=0\, ,
\ee
and hence $H_{abc}=0$ except for $D=6$ case. However, since in additoin all the spatial components of the Ricci tensor vanish, this result follows even in $D=6$ case.

Therefore, the only non-vanishing component of
$H_{\m\n\l}$ are  $H_{uab}(u,\BM{x})$. This means that
$H_{\mu\nu\lambda}=l_{[\mu}P_{\nu\lambda]}$, so that the field
strength $H_{\mu\nu\lambda}$ is aligned to the null Killing vector
$l_{\mu}$.

Using (\ref{det}) one has
\be
H_{\m\nu\;\;\;;\l}^{\;\;\;\l}=H_{\m\n\;\;\;,\l}^{\;\;\;\l}\, ,
\ee
and the field equations (\ref{5})--(\ref{6}) reduce to
\bea\label{rf1}
{ }\hspace{-0.5cm}(a_u)_{,a}^{,a}-\partial_u(a_a^{\;,a})
&=&{1\over4}\left( f_{ab}f^{ab}-H_{uab}H_u^{\;\;ab}\right) + \kappa
p_u\, ,
\\\label{feom}
f_{ab}^{\;\;\;,b}&=&-\kappa p_a\, ,
\\
\label{rf3} H_{ua\;\;,b}^{\;\;\;b}&=&8\kappa J_{ua}\, .
\eea
The last two relations are linear differential equation in  the
$n-$dimensional Euclidean space $(n=D-2)$. They can be solved for
$f_{ab}$ and $H_{uab}$ once the source $J_{ua}$ and the distribution
for the source for gravito-magnetic field $f_{ab}$ is given. After
this we can solve the first equation for $a_u$, which for a given
right-hand-side is also linear.

On the other hand, we can also solve the constraint $H_{abc}=0$ by
the following ansatz for the 2-form potential
\be
B_{\m\n}=A_{\m}l_{\n}-A_{\n}l_{\m}\;.
\ee
 From $l^{\m} B_{\m\n}=0$, we have
\be\label{consa}
l^{\mu} A_{\mu}=0.
\ee
This is equivalent to choose a gauge so that the only non-vanishing
component of $B_{\m\n}$ is $B_{ua}=A_a(u,\BM{x})$, and of
$H_{\m\n\l}$ is
\be\label{Huab}
H_{uab}=\partial_b A_a-\partial_a A_b\equiv F_{ba}.
\ee

Note that the constraints (\ref{cons}) and (\ref{consa}) are preserved
by the gauge transformation
\be\label{gaugea}
A_{\m} \rightarrow A_{\m}+\partial_{\m} \Lambda(u,\BM{x}).
\ee

Let us denote
\bea
\Phi &=& 2a_u\hhh\BM{a}=a_a\hhh\BM{f}=f_{ab}\, \\
\BM{A} &=&A_a\hhh\BM{F}=F_{ab}\hhh\BM{J}=J_{ua}\hhh \BM{p}=p_a\, .
\eea
In these notations the gyraton metric is
\be
ds^2=d\bar{s}^2+\Phi du^2 + 2(\BM{a},d\BM{x})du\, ,
\ee
and the field equations (\ref{rf1}) and (\ref{rf3}) reduce to
\be\label{Phi}
  \Delta \Phi -2\partial_u(\BM{\nabla} \cdot \BM{a})
  ={1\over2}\left( \BM{f}^2-\BM{F}^2 \right)+2\kappa p_u\, ,
\ee
\be\label{eoma}
\Delta \BM{A}+\BM{\nabla}(\BM{\nabla} \cdot \BM{A})=8\kappa \BM{J}.
\ee
Here $\BM{
\nabla}=\partial_a$, and $\Delta$ is the Laplacian
operator in the $n-$dimensional Euclidean space.

Using the coordinate, (\ref{gaugg}), and electromagnetic,
(\ref{gaugea}), gauge transformations one can put
\bea
&&\nabla \cdot \BM{A}=0\, ,
\\
&&\nabla \cdot \BM{a}=0\, .
\eea
For these gauge fixing conditions the equations (\ref{Phi}), (\ref{feom}) and
(\ref{eoma}) take the form
\bea\label{P}
\Delta \Phi&=&{1\over2}\left( \BM{f}^2-\BM{F}^2 \right)+2\kappa
p_u\,
,\\
\label{a}
\Delta \BM{a}&=&\kappa \BM{p}\, ,\\
\label{A} \Delta \BM{A}&=&8\kappa \BM{J}\, .
\eea
It is interesting to note that the magnetic and gravitomagnetic terms
enter the right hand side of (\ref{P}) with the opposite signs. A
special type of solutions is the case when these terms cancel one
another, so that the equation for $\Phi$ outside the matter source
becomes homogeneous. We call such solutions {\em saturated}. The
condition of saturation is
\be
\BM{f}^2=\BM{F}^2
\ee
which can be achieved by letting $\BM{p}=8 \BM{J}$ as suggested by
(\ref{a}) and (\ref{A}).

\section{Ring String Gyratons}

\subsection{Green functions}

The system of equations (56)--(58) is well defined for distributed
sources $p_u$, ${\bf p}$, and ${\bf J}$. In a general case, solutions for the equations \eq{P}--\eq{A} can be
obtained by using the Green function of the Laplace operator. A
solution of the equation
\be\label{J}
\Delta \psi= j(u,\BM{x})
\ee
is
\be\label{psi}
\psi(u,\BM{ x})= -\int d\BM{ x}' {\cal G}_n(\BM{ x},{\BM{
x}'}) j(u,\BM{ x}')\, .
\ee
Here ${\cal G}_n(\BM{ x},{\BM{ x}'})$ is the Green's function for the
$n-$dimensional Laplace operator
\be\label{gf}
\Delta {\cal G}_n(\BM{ x},{\BM{ x}'})
=-\delta (\BM{ x}-{\BM{ x}'})\, ,
\ee
which can be written in the following explicit form
\be\label{gf2}
{\cal G}_2(\BM{ x},{\BM{ x}'})=-{1\over 2\pi}\ln |\BM{ x}-\BM{ x}'|\, ,
\ee
\be\label{gfn}
{\cal G}_n(\BM{ x},\BM{ x}')=
{g_n \over |\BM{ x}-\BM{ x}'|^{n-2}}
\hhh n>2 \, ,
\ee
where $g_n=1/[(n-2){\cal A}_n]$ and ${\cal A}_n$ is given by
\eq{ss}. It should be emphasized that the retarded time $u$ plays
the role of the external parameter if a source term depends on it.
The dependence of the field on $u$ can be obtained by solving
"static" equations. After one obtains a solution of static
equations one can simply make the coefficients which enter the
solution to be $u-$dependent.

Since the equations \eq{a} and \eq{A} are linear, one can solve them for the limiting case
when the size of the source tends to zero. However, because the presence of the terms quadratic in ${\bf F}$ and ${\bf f}$ in the right hand side of equation \eq{P}, its solution by the transition to the point like limit of sources may  be not uniquely determined and  depend on detailed
structure of the sources. In what follows we restrict ourselves by
considering the saturated solutions where this problem does not
occur. We also assume that the source of the gyraton field is a string which has a
form of a ring and is moving with the speed of light in the direction
orthogonal to the plane of the ring. Let us consider this case in
more details.

\subsection{KR field of a ring string}

We choose coordinates $\BM{x}$ in the transverse to the motion plane
in such a way that the string ring is located in the
$(x^3$-$x^4)-$2-plane, and denote the coordinates in the orthogonal
subspace by $\BM{y}=(y^A)$, \ $A=5\ldots D$. Let $(\rho, \theta)$ be
polar coordinates in the string 2-plane
\be
x^3=\rho \cos \theta \hh x^4=\rho \sin \theta\, .
\ee
The ring-string worldsheet is defined by the following equations
\be
v=\tau\hhh u=u_0\hhh X^3=\rho_0\cos\sigma \hhh X^4=\rho_0 \sin\sigma\,
,
\ee
and $X^A=0$. The non-vanishing
component of the current $J^{\mu\nu}$ is $\BM{J}=J^{v a}$
\bea
\BM{J}&=& {1\over 2}q\upsilon(u)\BM{I}\hh
\BM{I}=I_{\hat{\theta}}\BFE{\theta}\, ,\\
\BFE{\theta}&=&- \sin\theta \, \BFE{3}+ \cos\theta \, \BFE{4}\, ,
\\
I_{\hat{\theta}}&=&{\delta(\rho-\rho_0)\over \rho_0}
\delta^{D-4}(\BM{y})\, .
\eea
Here $\BFE{\theta}$ is a unit vector in the $\theta-$direction,
$\BFE{a}$ are unit vectors along the axes $x^a$, and $\upsilon(u)=
\delta(u-u_0)$. If the string source is not localized at $u=u_0$ but
is smeared in time, the function $\upsilon(u)$ is smooth. We assume
that $\upsilon(u)$ is normalized so that $\int du \upsilon(u)=1$.

The solution of the equation (\ref{A}) is \cite{rem}
\be\label{AD}
\BM{A}(u,\BM{x})= \left\{
\begin{array}{ll}
\displaystyle{ -{1\over2} g_{D-2}Q \upsilon(u) \int {d\BM{x}'
\BM{I}(\BM{x}')\over |\BM{x}-\BM{x}'|^{D-4}}}\, ,
& \mbox{for  }D>4\, ;\\
& \\
\displaystyle{{1 \over 4\pi} Q \upsilon(u) \int d\BM{x}'
\BM{I}(\BM{x}')\ln |\BM{x}-\BM{x}'| }\, ,& \mbox{for  }D=4\, .
\end{array}
\right.
\ee
Here the charge $Q$ is given by (\ref{charge}) and
\be
|\BM{x}-\BM{x}'|=[\lambda +\rho^2+{\rho'}^2-2\rho\rho'
\cos(\theta-\theta')]^{1/2}\, ,
\ee
and  $\lambda=|\BM{y}-\BM{y}'|^2$ (for $D>4$) and $\lambda=0$ (for
$D=4$).

Since the geometry is cylindrically symmetric, it is sufficient to
calculate $\BM{A}$ at $\theta=0$. Since the azimuthal integrations
in (\ref{AD}) is symmetric about $\theta'=0$, the component of the
current in the direction $\BM{e}_{3}$ does not contribute. This
leaves the only component in the direction of $\BM{e}_{4}$, which is
$A_{\hat{\theta}}$. Since $\BM{e}_{\hat{\theta}}=\rho
\partial_{\theta}$ one has $A_{\theta}=\rho^{-1}A_{\hat{\theta}}$. Thus
we have
\bea
A_{\theta}&=&-{1\over 2\rho} g_{D-2}Q \upsilon(u)  {\cal B}_D\,
,\mbox{ for  }D>4;
\\
\label{int4} A_{\theta}&=&\displaystyle{{1 \over 4 \pi\rho} Q
\upsilon(u) \int_0^{2\pi} d\theta' \cos \theta'\, \ln F_0}\, , \
D=4\, ,
\eea
where
\bea
F_{\lambda}&=&\lambda +\rho^2+{\rho_0}^2-2\rho\rho_0
\cos(\theta')\hh \lambda=\BM{
y}^2\, ,\\
{\cal B}_D&=& \int_{0}^{2\pi} {d\theta'\, \cos \theta'\over
F_{\lambda}^{(D-4)/2}}\, .
\eea

\subsubsection{4D case}

In the 4-dimensional case, $D=4$, the integral \eq{int4} can be
easily taken and the answer is
\be\label{d4a}
A_{\theta}=-{Q\upsilon(u)\over 2 \rho_0}
\left[\Theta(\rho_0-\rho)+(\rho_0/\rho)^2\Theta(\rho-\rho_0)
\right]\, ,
\ee
where $\Theta$ is the Heaviside step function.  In this case the only
non-vanishing component of the 3-form flux
\be\label{d4aH}
H_{u\theta\rho}=A_{\theta,\rho}= Q \upsilon(u) {\rho_0\over
\rho^3}\Theta(\rho-\rho_0)\, .
\ee
The flux vanishes for $\rho<\rho_0$ and it is discontinuous at
$\rho=\rho_0$. However, the jump of the flux value at $\rho=\rho_0$
is {\it finite} and proportional to the charge $Q$.

\subsubsection{Even dimensional case}

For $D>4$ we consider first the case when $D$ is even.
We put $D=2m+6$, then ${\cal B}_D=H_m$, where
\be\label{de4a}
H_m= \int_{0}^{2\pi} {d\theta'\, \cos \theta'\over
F_{\lambda}^{m+1}} \, .
\ee
The calculations give
\be
H_0={\pi \over \rho\rho_0}\left[ {\lambda +\rho^2+\rho_0^2\over
\sqrt{ (\lambda +(\rho+\rho_0)^2)(\lambda
+(\rho-\rho_0)^2)}}-1\right]\, .
\ee
For $m>0$ one has
\be
H_m ={(-1)^m\over m!}{d^{m}H_0\over d\lambda^m}\, .
\ee
Note that the above solutions are singular at the location of the
ring string, i.e., at $\rho=\rho_0$ and $\BM{y}=0$.

\subsubsection{Odd dimensional case}

Let $D=2m+5$, then ${\cal B}_D=J_m$, where
\be\label{do4a}
J_m= \int_{0}^{2\pi} {d\theta'\, \cos \theta'\over
F_{\lambda}^{m+1/2}} \, .
\ee
The calculations give
\be\label{J0}
J_0={2 \left[ (2-k^2) K(k)-2E(k)\right]\over k \sqrt{\rho\rho_0}}\,
,
\ee
\be\label{kdef}
k^2={4\rho\rho_0\over \lambda+(\rho+\rho_0)^2}\le 1\, .
\ee
Note that $k=1$ only at the location of the ring string, and $k<<1$
far away from the ring string or near the $\rho=0$ axis.

In the above $E(k)$ and $K(k)$ are the complete elliptic integrals
defined by
\bea\nn
&& K(k):=\int_0^{\pi/2} {dz \over \sqrt{1-k^2\sin^2 z}}\;,
\\\nn
&& E(k):=\int_0^{\pi/2} dz \sqrt{1-k^2\sin^2 z}.
\eea
Note that $E(k)$ is a monotonically decreasing function from
$E(0)=\pi/2$ to $E(0)=1$, and $K(k)$ is a monotonically increasing
function from $K(0)=\pi/2$ to $K(1)=\infty$. Note that $K(k)\sim
\ln(1-k)$ as $k \rightarrow 1$.

For $m>0$ one has
\be\label{Jm}
J_m=(-1)^{m}{2^{m}\over (2m-1)!!} {d^{m}J_0\over d\lambda^m}\, .
\ee
The complete elliptic integrals possess the properties
\bea\label{KK}
{dK(k)\over dk}&=&{E(k)\over k(1-k^2)}-{K(k)\over k}\, ,
\\
\label{EE}
{dE(k)\over dk}&=&{E(k)-K(k)\over k}\, .
\eea
For this reason the expression $J_m$ for any $m$ has the same structure
\be
J_m= A_m K(k)+B_m E(k)\, ,
\ee
where $A_m$ and $B_m$ are some algebraical functions of $k$ and
$\rho\rho_0$, which can be found by using \eq{KK}--\eq{EE}.

   Note that the above solutions are singular on the location of
the ring string due to the divergence of $K(k)$ at $k=1$.

\subsection{Gravito-magnetic potential $\BM{a}$}

Since the equation for the gravito-magnetic potential $\BM{a}$,
\eq{a}, differs from the \eq{A} only by a constant coefficient one
can directly obtain a solution for $\BM{a}$ from $\BM{A}$. For the
source $\BM{p}=j(u) \BM{I}$ it is sufficient to make the following
substitution $4q\upsilon(u)\to j(u)$ in the expression for $\BM{A}$.
For saturated solutions we shall further require $j(u)=
4q\upsilon(u)$.

\subsection{Potential $\Phi$}

The equation \eq{P} for $\Phi$ is linear. Let us write its solution in
the form
\be
\Phi=\varphi+\psi\, ,
\ee
where
\be
\Delta \psi=2 \kappa p_u\, .
\ee
Assuming that the mater source is localized on the ring string, one
has
\be
p_u=\varepsilon(u) I_{\hat{\theta}}\, .
\ee
The corresponding solution is
\be
\psi(u,\BM{x})=
\left\{
\begin{array}{ll}
&\displaystyle{-2g_{D-2}  \kappa \varepsilon(u) {\cal C}_D\, ,\mbox{for  }D>4}\, ;\\
&\\
&\displaystyle{{ \kappa \varepsilon(u)\over \pi}  {\cal C}_4\, ,
\mbox{for  }D=4}\, .
\end{array}
\right.
\ee
Here
\bea
\label{CD}
{\cal C}_{D>4}&=& \int_{0}^{2\pi} {d\theta'\, \over F_{\lambda}^{(D-4)/2}}\, ;\\
\label{C4} {\cal C}_4&=&   \int d\theta' \, \ln F_0\,  .
\eea

\subsubsection{4D case}

In the 4-dimensional case,  the integral \eq{C4} can be easily taken
and the answer is
\be\label{d4p}
{\cal C}_4=-4\pi \left\{
\begin{array}{ll}
\displaystyle{\ln \rho}\, ,& \mbox{for  }\rho>\rho_0\, ;\\
\displaystyle{\ln \rho_0}\, ,& \mbox{for  }\rho<\rho_0\, .
\end{array}
\right.
\ee

\subsubsection{Even dimensional case}

For $D>4$ we consider first the case when $D$ is even.
We put $D=2m+6$, then ${\cal C}_D=N_m$,
\be\label{de4p}
N_m=\int_{0}^{2\pi} {d\theta'\, \over F_{\lambda}^{m+1}} \, .
\ee
The calculations give
\be
N_0= {2\pi \over \sqrt{ (\lambda +(\rho+\rho_0)^2)(\lambda
+(\rho-\rho_0)^2)}}\, .
\ee
For $m>0$ one has
\be
N_m ={(-1)^m\over m!}{d^{m}N_0\over d\lambda^m}\, .
\ee

\subsubsection{Odd dimensional case}

Let $D=2m+5$,
then ${\cal C}_D=L_m$,
\be\label{do4p}
L_m= \int_{0}^{2\pi} {d\theta'\, \over F_{\lambda}^{m+1/2}} \, .
\ee
The calculations give
\be
L_0= {2 k K(k)\over \sqrt{\rho\rho_0}}\, ,
\ee
where $k$ is given by \eq{kdef}.

 For $m>0$ one has
\be\label{Lm}
L_m=(-1)^{m}{2^{m}\over (2m-1)!!} {d^{m}L_0\over d\lambda^m}\, .
\ee
Using this relation and relations (\ref{KK}) and (\ref{EE})
it is possible to write $L_m$ as
\be
L_m= C_m K(k)+D_m E(k)\, ,
\ee
where $C_m$ and $D_m$ are some algebraical functions.

For the saturated ring-string solutions $\varphi=0$, so that
$\Phi=\psi$. As for the KR field, the higher dimensional ($D>4$)
solutions of $\Phi$ are singular at the location of the ring string.
The obtained relations in this section  allow one to write the
gyraton metric (\ref{mansatz}) and the 3-form flux (\ref{Huab}) in
an explicit form. The solutions contain 2 arbitrary function of $u$,
$\upsilon(u)$ and $\varepsilon(u)$, which are related to the angular
momentum (string current) and energy density of the ring-string
source.  For non-saturated case, one needs in addition to obtain a
solution of the equation
\be
\Delta \varphi={1\over 2}(\BM{f}^2-\BM{F}^2)\,
\ee
with an  explicitly known right hand side. The relation \eq{psi} gives
the integral representation for the solution.

\section{Conclusions}

In general it is difficult to obtain  analytically in an explicit
form a solution of the Einstein equations for a moving extended
graitating object because of the non-linearity the equations. In this
paper we achieve the goal by constructing the full supergravity
solution due to a boosted closed string coupled to the Kalb-Ramond
field. We basically generalized the method of constructing of
solutions for the point charged gyraton moving with the speed of
light \cite{FrZe:05a} to the string case. The gyraton metric has the
special property that all the curvature invariants constructed from
the curvature and its covariant derivatives vanish
\cite{FrFu:05,FrIsZe:05}. As a pecial case, we considered a ring
string source moving with the speed of light in a $D$-dimensional
spacetime.  The full metric and the KR field due to its back reaction
are analytically constructed, and the quantities in the solutions
related to angular momentum and energy density are identified. The
configurations are singular at the location of the ring string source
for $D>4$ cases but not for $D=4$. Physically, our special  ring
string solution is the theoretical realization of the boosted closed
string produced in the high energy experiments, and the
time-dependent background variation could, in principle, be detected
by the gravitational wave interferometers.

  The generalization of our construction to the $p$-brane gyratons is
straightforward, especially the $3$-brane gyraton of the vacuum
supergravity solution is the generalization of the pp-wave background
\cite{bmn,blau}. However, the explicit solutions corresponding to the
the boosted $p$-brane of special shape are  model dependent.
It will be interesting to examine the string theory in the vacuum
$3$-brane gyraton background and its dual picture in Yang-Mills
theory. Finally, we would like to mention that the solutions found in \cite{krtous1,krtous2}
by boosting the black ring are related to the solution  found in this paper. It will be also interesting to 
explore the connection.

\noindent
\section*{Acknowledgments}
\noindent

The research was supported in part by US National Science Foundation
under Grant No. PHY99-07949 and by Taiwan's NSC grant
94-2112-M-003-014. One of the authors (V.F.) is grateful for
hospitality to KITP, Santa Barbara, and to the Depatment of Physics,
National Taiwan Normal University, Taipei.


\end{document}